# Large Ca Isotope Effect in $CaC_6$

D.G. Hinks, D. Rosenmann, H. Claus, M.S. Bailey, and J.D. Jorgensen

Materials Science Division, Argonne National Laboratory, Argonne IL 60439

ABSTRACT

We have measured the Ca isotope effect in the newly discovered superconductor $CaC_6$. The isotope effect coefficient, $\alpha$, is 0.50(7). If one assumes that this material is a conventional electron-phonon coupled superconductor, this result shows that the superconductivity is dominated by coupling of the electrons by Ca phonon modes and that C phonons contribute very little. Thus, in contrast to $MgB_2$, where phonons in the B layers are responsible for the superconductivity, in $CaC_6$ the phonons are primarily modes of the intercalated Ca.

The recent discovery of superconductivity [1,2] at 11.5K in a Ca graphite intercalation compound (GIC) has renewed interest in the superconducting behavior of this class of materials. Superconductivity in the alkali GICs was first reported in 1965 [3] however $T_c$s were well below 1K. The low $T_c$s, the reactivity to air, difficult synthesis and irreproducible superconducting transitions have been impediments to many workers. As an example, the original $T_c$ reported by Hannay et al. [3] was 390 mK for $KC_8$ however initial attempts to reproduce this result were elusive. Poitrenaud [4] could find no evidence for superconductivity down to 300 mK. A flurry of activity around 1980 [5-7] showed that the $T_c$ of the $KC_8$ compound was much lower and could vary over a considerable range (128 to 198 mK) depending on synthesis conditions [7].

The first synthesis of $CaC_6$ was done by Weller et al. [1] using Ca vapor transport. These samples were not homogeneous and the resulting superconducting transitions were broad. Emery et al. [2] was able to synthesize bulk samples (at least 60% phase fraction) with sharp transitions using a flux. The highly oriented pyrolytic graphite (HOPG) was immersed in a Li-Ca solution (75 atomic% Li) at approximately 350°C for several days. Li is initially intercalated into the HOPG followed by a slow exchange of Ca for Li [8]. Using lower temperatures and different Li-Ca flux compositions, ternary Li-Ca-C compounds can also be obtained [9,10]. One of these materials, $Li_3Ca_2C_6$, was shown to have a superconducting onset at 11.15K [10]. The hypothesis concerning this ternary GIC is that the Li and Ca intercalate into separate layers and that the superconductivity comes from the domains intercalated with Ca.

The nature of superconductivity in GICs is still an area of some uncertainty. In all the donor GICs (alkali metals, alkaline earth metals, rare earths, etc.) electrons are transferred from the donor atom to the graphene π bands. Two bands are important for superconductivity in the GICs, the 2D π bands and the 3D donor, or interlayer, band [11,12]. However, the nature of the paring mechanism is still not fully understood. In general the coupling of the phonon modes of the graphene layers to the π bands is weak with no coupling of the out-of-plane lattice modes and only weak coupling of the in plane modes. Assuming $T_c$ is determined by C phonons, the low $T_c$s of these materials might be expected based on the small coupling to the π electrons. The appearance of the high $T_C$ in $CaC_6$ is thus surprising. Csányi et al. [13] suggests that an electronic mechanism may be responsible for the high $T_c$. Mazin [14], on the other hand, suggests most of the coupling is due to Ca phonons. He assumes that the difference in $T_C$ between $CaC_6$ and $YbC_6$ is due mainly to the difference in mass. Based on this idea, he predicts a Ca isotope effect coefficient (IEC) of about 0.4. Calandra et al. [15], using density functional theory, calculated that the superconductivity is mediated somewhat equally by the Ca in plane and C out-of-plane vibrations with small contributions from other vibrations. Their calculated Ca and C isotope effect coefficients are 0.24 and 0.26, respectively.

The isotope effect coefficient (α) for an ion of mass M is defined as

$$\alpha = - d \ln T_c / d \ln M.$$

The Ca isotope effect coefficient is important for determining the contribution of the Ca phonon modes to superconductivity. The isotope effect coefficient has not been measured for any GIC to date, probably due to the low $T_c$s of the materials and the difficulty in obtaining reproducible superconducting transitions. $CaC_6$ is a material that shows a narrow, reproducible $T_c$ that is high enough to yield a large $\Delta T_c$ for isotopic substitution of either Ca or C. Normally for sp materials the cononical value of of α is 0.5, whereas d-electron systems usually show a reduced isotope effect coefficient [16]. For a multicomponent system the sum over the individual isotope effect coefficients should have a maximum value of 0.5. Thus, a $\Delta T_c$ of about 0.5K would be expected between $^{40}Ca$ and $^{44}Ca$ assuming a full Ca IEC. We have measured the Ca isotope effect coefficient using natural abundance Ca (Rare Earth Products, 99.98% purity) and $^{44}Ca$ (Oak Ridge 98.78 %, calculated atomic weight 43.94) intercalated into GE Advance Ceramics grade ZYA HOPG.

We have tried both flux intercalation, using 75 atomic% Li solutions, and vapor transport to prepare the samples. We find that vapor transport, while not in general giving fully intercalated samples, yields much sharper and reproducible superconducting transitions. The data in this paper are for vapor transport samples only. The HOPG was cut to a 2x2x2 $mm^3$ sample size. This sample was cleaved into 4 pieces with thickness of about 0.4mm. Thus these 4 samples were all taken from the same place in the larger

12x12x2mm$^3$ HOPG plate hopefully insuring the same chemical and physical properties. About 60mg of either natural abundance Ca ($^n$Ca) or $^{44}$Ca along with 2 HOPG samples were placed in each of two 40mm long by 9.5mm diameter stainless steel tubes. The tube ends were only squeezed shut to allow the evacuation of the tubes. The loading of the stainless tubes was done in a N$_2$ filled glove bag. The stainless tubes were transferred to a vacuum system and pumped to 4x10$^{-3}$ MPa. After 10 hr the temperature was raised to 450°C and held for 4 days. Pumping, to remove outgasing products, was continued for the entire time to keep the pressure low and the mean-free path of the Ca vapor large in the stainless tubes to ensure good Ca transport.

Magnetization measurements were performed in a non-commercial SQUID magnetometer [17]. The Earth's magnetic field is shielded by a µ-metal shield yielding a remnant magnetic field of less than 10 mGauss. Magnetic fields up to 50 Gauss can be obtained with a Cu solenoid.

Figure 1 shows the superconducting transitions for the 4 samples. The samples were cooled in a magnetic field of 1 Gauss (parallel to the c-axis of the HOPG) to 4.2 K and magnetization was then recorded on warming. Figure 1 shows the normalized magnetizations, the actual magnitude of the Meissner magnetization at low temperatures ranges from 4 to 8x10$^{-6}$ emu for the various samples. Note that the two samples of each isotope are very similar indicating that reproducibility of the intercalation is not a problem. The transition temperature width, using 10 to 90% criterion and averaging the duplicate samples, is slightly larger for the $^{44}$Ca sample, 0.46(3) vs. 0.37(3)K for the $^{44}$Ca and $^n$Ca samples, respectively.

It should be pointed out that the samples are not fully intercalated after 4 days of heating. This becomes obvious when comparing the field cooled (Meissner) magnetization with the zero-field cooled (shielding) magnetization. Figure 2 shows an example for one of the two $^{44}$Ca samples. The other 3 samples show similar behavior. The magnetic field is 1 Gauss and is parallel to the c-axis of the HOPG. At first glance it appears that there are two different transitions in the zero-field cooled state but that only the one with a higher T$_C$ displays a Meissner effect. However, this is not the case, instead, the zero-field cooled magnetization resembles that of a superconducting ring with a weak link [18]. The HOPG probably contains grain boundaries since it grows in a columnar fashion. These can become weak links with reduced critical current after Ca-doping. The two insets in Figure 2 demonstrate the schematic current flow in a ring with a single grain boundary (GB). For the zero-field cooled state, the induced shielding current at low temperatures flows along the outer perimeter of the sample. The induced current is smaller than I$_C$(GB) the critical current of the grain boundary (inset A). As the temperature is increased, the critical current of the grain boundary decreases monotonically. Eventually it will drop below the shielding current and the magnetic moment will decrease accordingly with increasing temperature as the shielding current is shunted to the inside of the annulus. When the grain boundary critical current reaches 0, which happens at about 10.7K for the sample in Figure 2, the current flow is schematically represented by inset B. The current through the grain boundary is zero. The bulk of the sample (annulus), however, is still superconducting and will show a transition at $T_c$ as the shielding currents go to zero. If more than one grain boundary is present, only the weakest one will have an effect on the zero-field cooled magnetization [18]. Thus the upper transition measures the true superconducting transition and the lower transition is an indirect measure of the grain boundary critical current.

This model naturally explains why in the field-cooled magnetization only the real transition is seen. During cooling of a ring through the superconducting transition flux will be expelled to the outside of the ring as well as into the bore, the current distribution in the field-cooled state being essentially the same as in inset B of Figure 2 (the currents flowing parallel to the grain boundary have no effect in the magnetic moment). If we make a radial cut through the superconducting shell, the circulating current would be interrupted and the shielding current will flow at all temperatures below $T_c$ as indicated in inset B of figure 2. In that situation the shielding curve would look similar to the field-cooled magnetization, just like the ten-fold enhanced field-cooled magnetization in figure 2. This exact behavior was previously seen for a ring sample with a single grain boundary in Reference 18.

Using the averaged extrapolated onsets of 11.64(3)K and 11.06(3)K for $^n$Ca and $^{44}$Ca, respectively, we obtain $\alpha(Ca) = 0.56(4)$. This isotope effect coefficient is at the BCS limit of 0.5 even without taking into account the contribution that C would make to the total IEC. Considering that this material is a sp system a value of 0.5 for the total $\alpha$ might be expected. However, it may be somewhat unreasonable to expect the C isotope effect coefficient to be negligible.

The value of the IEC will depend to some extent on the criterion for the determination of $T_c$ for the samples. This is important for this experiment since the measured transitions for the $^{44}$Ca samples are slightly broader and have a more rounded onset temperature. The reason for this is not known with any certainty however instability of the samples in ambient air and a possible impurity effect cannot be ruled out. Impurity effects could occur since the impurity contents of both Ca isotopes are very different. The major detected impurities in each of the Ca sources used are listed in reference 20. If one of these impurities was doping the $CaC_6$ and lowering the transition temperature, the onset temperature might be a better representation of $T_c$. The onset temperatures are 11.65(3)K and 11.20(3)K for $^n$Ca and $^{44}$Ca, respectively. The Ca isotope effect coefficient is then calculated to be 0.43(4).

Thus, the error in this measurement depends more on the definition of $T_c$ than on the temperature measurement; we find that the Ca isotope effect coefficient is between 0.43 and 0.56 depending on the definition of $T_c$. The average is $\alpha = 0.50(7)$. Further effort will be required to determine it to better accuracy.

It is instructive to compare two seemingly similar materials, $MgB_2$ and $CaC_6$. Structurally these materials are similar, containing 2D hexagonal nets of B or C separated by an alkaline earth ion. In $MgB_2$ charge transfer from the Mg to the B network is complete. The interlayer band lies above the Fermi level and is unimportant in this material. It is charge transfer from the B σ bands to the B π bands that create holes in the σ bands which host the strongest superconducting interaction. These carriers couple strongly to the in-plane B vibrations leading to high-$T_C$ superconductivity. In this material, B shows a large isotope effect (0.30) whereas Mg shows an isotope effect of only 0.02 [21]. In $CaC_6$ charge transfer is not complete leading to an interlayer (or intercalate) band at the Fermi level. This band couples into the Ca phonons and, through hybridization with the π bands, is able to enhance the coupling to the C vibrations. The degree of hybridization of these bands will control the isotope effect coefficient of Ca and C. We find a very large Ca isotope effect and, consistent with the BCS limit of 0.5, would expect little in the way of coupling to the C phonons.

Regardless of the possible errors in this experiment, Ca clearly supplies a large phonon contribution for superconductivity in $CaC_6$. It would be instructive to measure the C isotope effect coefficient and refine the value of the Ca isotope effect coefficient to see if the total isotope effect is possibly greater than 0.5. If so, this would then be the second example of a total isotope coefficient greater then 0.5, underdoped high $T_C$ copper oxides being the first [22]. In general our results are consistent with the work of Mazin [14]. He deduces a value of 0.4 for the Ca isotope effect coefficient close to the lower limit of our value. Calandra et al. [15] calculated that the isotope effect coefficients for Ca and C are about equal but these values are not consistent with our result indicating that their estimate of the hybridization of the intercalate band with the π bands may be too large.

This example of the intercalated cations, rather than the 2D graphite layers, being responsible for superconductivity suggests an additional strategy for searching for new superconducting materials. Perhaps one should look for systems where a 2D template can be used to "suspend" or arrange light atoms in such a way that phonon amplitudes are unusually large, owing to longer that normal distances between the cations and giving rise to strong electron-phonon coupling.

ACKNOWLEGEMENTS
This work was supported by the U. S. Department of Energy, Division of Basic Energy Sciences -- Materials Sciences, under contract W-31-109-ENG-38. We wish to thank Ken Gray for useful discussions about the interpretation of these results.

Figure Captions

Figure 1. Normalized magnetization versus temperature for two samples with natural abundance Ca ($^n$Ca) and two samples with the $^{44}$Ca isotope. The data were taken on warming after field cooling in 1 G. The two samples represented by the solid symbols were measured within hours after removing from them from the furnace; the other two (open symbols) were measured 2 days later after storage in a $N_2$ filled dry bag.

Figure 2. Field cooled and zero-field cooled magnetization in 1G for one of the $^{44}$Ca sample (solid triangles in Fig. 1). Also shown are the field cooled data scaled up by a factor of 10. Inset A shows the shielding current path when the grain boundary is fully conducting and inset B shows the current path when the grain boundary becomes insulating (see text).

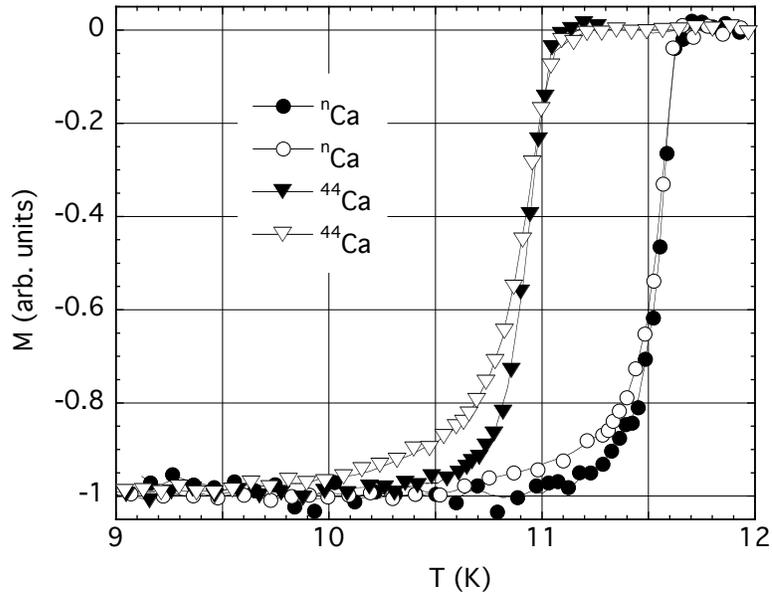

Figure 1

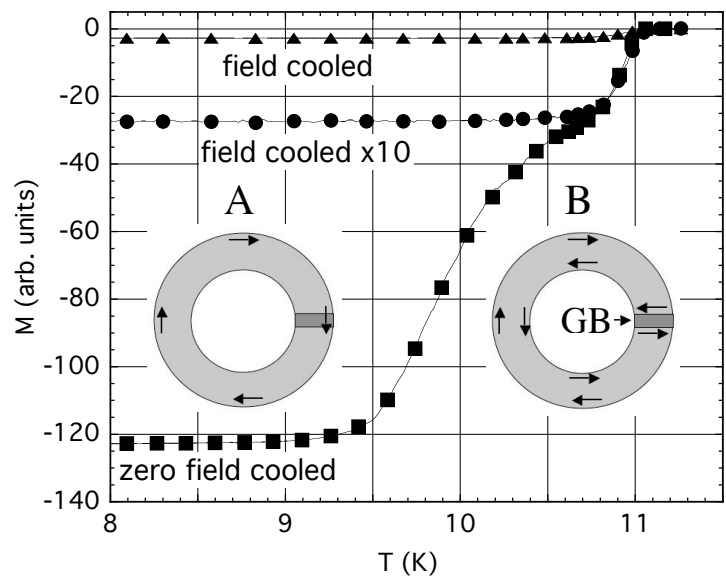

Figure 2